# Monitorization of the H–O Bond Flexibility

Chang Q Sun[1,2,*], Chunyang Nie[1], Yongli Huang[3], Yong Zhou[1], Lei Zhang[4,*], Biao Wang[1,5,*]

**Abstract**

**Unlike conventional thought, the H–O bond is flexible, instead, and sensitive to perturbation. Introducing an index *m* to correlate the H–O bond energy $E_H$ with its length $d_H$ in the form of $E_H = d_H^{-m}$ has enabled correlation, calibration, and quantification of the $d_H$, $E_H$, shift of vibration frequency $\Delta\omega_H$ (stiffness), and the O 1s energy $\Delta E_{1s}$, and the O:H nonbond length $d_L$, with ":" denoting the lone pair of oxygen. The known ($d_H$, $E_H$, $\omega_H$) values of (1.00, 3.97, 3200) for bulk water and (0.90 Å, 5.1 eV, 3610 cm$^{-1}$) for the dangling H–O bond resulted in the value of *m* = 2.3683. This exercise empowers the electron and phonon spectroscopies with the Tight-binding approach, enabling a referential database to synchronically quantize the relaxation and flexibility of these identities for water and substances involving the H–O bond during phonon spectroscopy.**

[1] Research Institute of Interdisciplinary Sciences (RISE) and School of Materials Science & Engineering, Dongguan University of Technology, Dongguan 523808, Guangdong, China (22022130@dgut.edu.cn; Chunyangnie@dgut.edu.cn)
[2] College of Engineering, Nanyang Technological University, Singapore 639798, Singapore
[3] School of Materials Science & Engineering, Xiangtan University, Xiangtan 411105, China (Huangyongli@xtu.edu.cn)
[4] School of Materials Science & Engineering, Beijing Institute of Technology, Beijing 100081, China; (zhanglei@bit.edu.cn)
[5] Guangdong Provincial Key Laboratory of Extreme Conditions, Dongguan 523803, China
* Corresponding authors: (wangbiao@mail.sysu.edu.cn, BW; zhanglei@bit.edu.cn, LZ; 2022130@dgut.edu.cn; CQ).

TOC

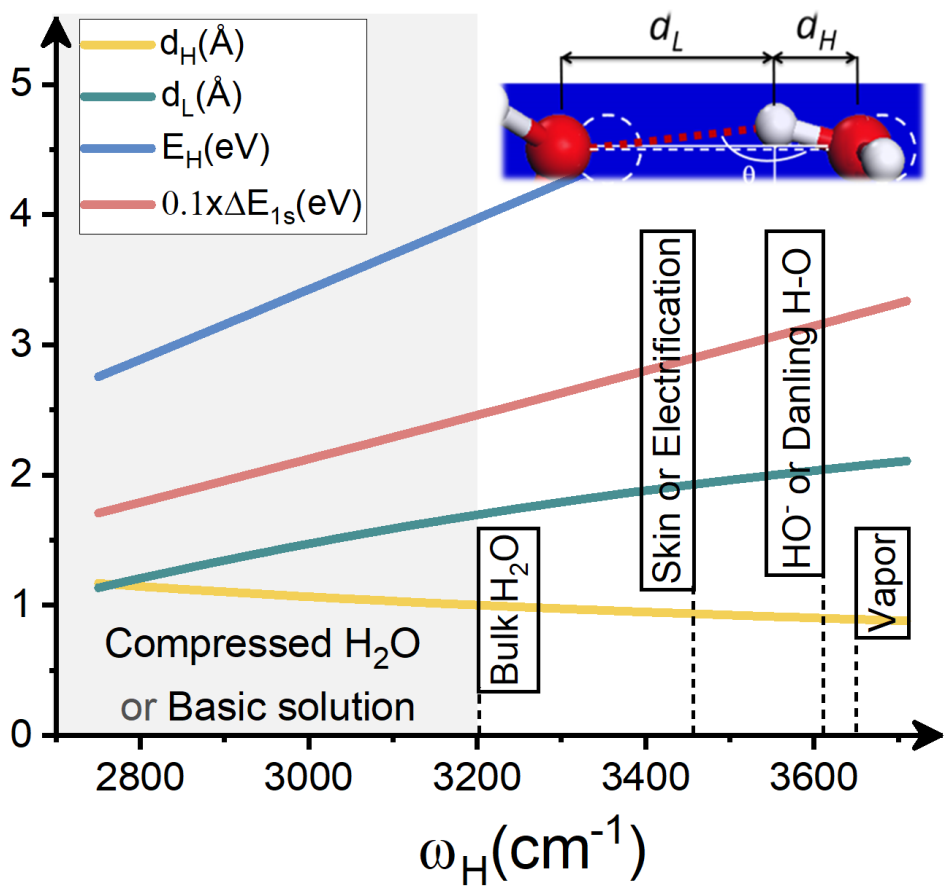

A referential database enabled the *in situ* quantification of H–O bond length-energy-stiffness, the O:H nonbond length, and the $O_{1s}$ level relaxation and flexibility during phonon spectroscopy.

Much focus has been on the intermolecular X···H attraction, defined as the hydrogen bond[1], isolating from the intra-molecular H–O polar-covalent bond that is often assumed rigid in various modeling systems for water [2-10]. For instance, the elegantly used TIP4Q/2005 model [11] consists of fixed four-point charges, constant H−O bond length, and unchangeable ∠H–O–H angle. The modeling employs a molecule as the primary structure unit to deal with the manner and mobility of molecular motion. The models feature well crystal structures of water [12], ice nucleation [13], liquid-solid transition [14], evaporation [15], and the lifetime of a certain molecule residing in a specific molecular site [16]. However, the description of phenomena such as ice regelation - melting point depression by compression[17] and Mpemba paradox - hot water fast cooling[18] are beyond the scope of molecular dynamics. Theoretical reproduction of anomalies such as density evolution during ice formation[19] is hardly feasible, though numerous potentials have been employed in density function theory and machine learning computations [20, 21].

Furthermore, concerns remain about what one can learn from electron and phonon spectroscopies and how the spectral data is properly digested and presented regarding substances involving the H–O bonds such as $M(OH)_n$, $H_2O_2$, $H_2O$, $HO^-$, $H_3O^+$, and aqueous solutions. The subscript n∈(1,3) denotes the valence value of a metal M such as Na, Mg, and Al. The H–O bond vibrates in the 2800-3700 cm$^{-1}$ frequency range depending on the coordination environment or applied perturbation such as pressure, temperature, electric field, or molecular undercoordination, regardless of its crystallizing constituents.

Recent progress demonstrates that the O—O repulsive coupling of the intermolecular O:H and the intramolecular H–O interactions is critical to the performance of the three-body coupling O:H–O bond and the functionality of water and ice [22]. The ":" denotes an electron lone pair of oxygen formed upon its $sp^3$-orbital hybridization that happens when its two bond orbitals are fully occupied [23]. As shown in **Figure 1**a and Figure A1, the engagement of the O—O repulsion results in the cooperative relaxation of the segmented O:H–O bond when perturbed, following the hydrogen bond cooperativity and polarizability (HBCP) regulation [22]. Mechanical compression or Liquid cooling shortens and stiffens the O:H nonbond while doing the stronger H−O bond contrastingly; however, quasi-solid (QS, an intermediate phase transiting Liquid to Ice-I) phase cooling, molecular undercoordination, or electrostatic polarization do the otherwise, associated with strong polarization. The O atoms of the O:H–O dislocate in the same direction but by different amounts, the O:H always relaxes more in extent than the H–O. The O:H–O bond cooperative relaxation is readily characterized using Raman scattering or infrared absorption spectroscopy with spectral signatures of $\omega_H$ centered at 3200 and $\omega_L$ at 200 cm$^{-1}$ for the bulk phase. Raman spectroscopy unveiled [24] that increasing pressure up to 40 GPa stiffens the O:H phonon of 20-K ice from 200 to 500 cm$^{-1}$ and lowers the H–O vibration frequency from 3350 to 2500 cm$^{-1}$, evidencing the essentiality of the O—O repulsive coupling interaction [25] that links the intra- and intermolecular interactions[12]. The introduction of the HBCP regulation has enabled resolutions to multiple issues regarding water irregularity [22], solvation interfacial bonding dynamic[26], and structural stability and energy storage of energetic explosives [27, 28].

We show herein that the zeroth (bond energy) and the second (curvature of the interatomic potential) Taylor coefficient of the H–O bond potential define intrinsically the shifts of the O 1s energy $\Delta E_{1s}$ and the H–O vibration frequency $\Delta \omega_H$, respectively. An introduction of the bond nature index $m$ is necessary to feature the H–O bond relaxation $\Delta d_H$ dependence of the $E_H$, $\Delta \omega_H$, $\Delta E_{1s}$, and the O:H distance $d_L$. One can thus readily quantify these parameters and their flexibility *in situ* from the presented referential database during phonon spectroscopy (See Table A1).

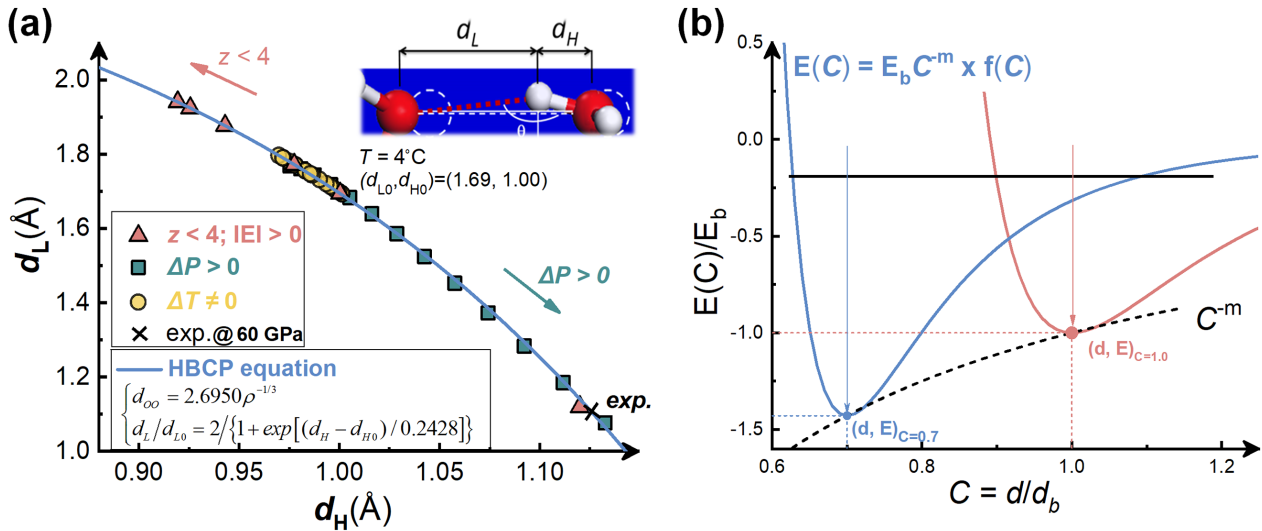

**Figure 1 | Illustration of the HBCP regulation and the bond nature index *m*.** (a) HBCP formulation (the inset equation) for the O:H nonbond ($d_L$) and H–O bond ($d_H$) length cooperativity under perturbation of compression P, heating T, electrification E ≠ 0, and atomic undercoordination $\Delta z < 0$ [22]. The $d_{OO}$ describes O—O distance and ρ the mass density of the tetrahedrally-coordinated water. Inset b illustrates the three-body O:H–O bond with H as the coordination origin. The H attracts the left-hand-sided O and

the O:⇔:O repulsion couples the inter- and intramolecular interactions. (b) The bond nature index $m$ correlates the potential depth $E$ with the relaxable bond length $d$. $C = d/d_b$ is the reduced bond length and $C^{-m} = E(C)/E_b$ is the reduced energy[29, 30] with subscript b denoting the bulk reference.

**Figure 1**b illustrates the essentiality of introducing the bond nature index $m$ for generalizing the two-body atomic potential. A perturbation shifts the equilibrium of the potential along the $C^{-m}$ curve rather than keeps it at a fixed position. Bond contraction absorbs energy while bond expansion does the inverse, which shifts the potential laterally in the f(C) form and vertically in the $C^{-m}$. The equilibrium coordinate $(C, C^{-m})$ calibrates the reduced bond length $C$ and energy $C^{-m}$, with subscript b representing the bulk reference ($C$ = 1). This correlation is universal to any two-body interactions with the $m$ value varying intrinsically from substance to substance [29, 30]. For carbon, $m = 2.56$; for Si, $m = 4.88$; for metals, it is 1.00.

The band theory of tight-binding approach[31] describes the performance of the interatomic bond and an electron moving in the v$^{th}$ orbit of an atom in the ideal bulk substance,

$$\begin{aligned} H &= H_0 + H' \\ \text{with} &\begin{cases} H_0 = -\dfrac{\hbar^2 \nabla^2}{2m} + U_{intra}(r) & \text{(Intra-atomic interaction of an isolated atom)} \\ H' = U_{inter}(r) & \text{(Inter-atomic interaction for the bulk)} \end{cases} \\ |v,i\rangle &\cong u_v(r+R_i)\exp(ikr) \quad \text{(Bloch wave-function)} \end{aligned}$$

(1)

The Hamiltonian for an isolated atom, $H_0$, is the sum of the kinetic energy and the intra-atomic potential energy experienced by an electron in a specific vth orbital. The eigen wavefunction, $|v,j\rangle$, is periodic in real space and meets the following orthogonal criterion with i and j denoting atomic positions:

$$\langle v,i|v,j\rangle = \delta_{ij} = \begin{cases} 1 & (i = i) \\ 0 & (i \neq j) \end{cases}$$

(2)

The energy of an electron in the vth orbital of an ideal bulk disperses with $k$ being the wavevector:

$$E_v(k) = E_{v0} + (\alpha_v + z\beta_v) + 2z\beta_v \Phi_v(k,R)$$

$$\text{with} \begin{cases} E_{v0} = -\langle v,i|H_0|v,i\rangle & \text{(Atomic core level energy)} \\ \alpha_v = -\langle v,i|H'|v,i\rangle \propto E_b & \text{(Exchange integral)} \\ \beta_v = -\langle v,i|H'|v,j\rangle \propto E_b & \text{(Overlap integral)} \end{cases} \quad (3)$$

The $H_0$ defines the referential energy level $E_{v0}$ of an isolated atom, from which the core band shifts when the $H'$ is involved or perturbed by an externally applied field. The $E_{v0}$ reduces its value from $10^3$ to $10^0$ eV until the vacuum level $E_0 = 0$ as the ν increases, or as one moves outwardly from the innermost orbit of an atom.

The core-level energy shifts from the reference $E_{v0}$ by an amount of $\Delta E_{vb}(C=1) = E_{vb} - E_{v0} = \alpha_v + z\beta_v$ upon $H'$ involvement during bulk formation. Further shift occurs once the dimer bond is relaxed under perturbation, such as mechanical or thermal loading, electrical polarization, and atomic/molecular under- or hetero-coordination. Meanwhile, perturbation turns the energy level into a band of $E_{vW} = 2z\beta_v\Phi_v(k,R)$ width with $\Phi_v$ being a crystal structure-sensitive function.

The exchange integral $\alpha_v$ and the overlap integral $\beta_v$ are proportional to the bond energy, $E_b$. Typically [31],

$$\Delta E_{vb} = \alpha_v + z\beta_v = \alpha_v(1 + z\,\beta_v/\alpha_v) \approx \alpha_v(1+3\%) \propto E_b. \quad (4)$$

A 3% contribution from the overlap integral to the $\Delta E_{vb}$ is negligibly small. For the deeper bands, this ratio is even smaller. Thus, the bulk energy shift $\Delta E_{vb}$ depends mainly on the bond energy $E_b$. Any relaxation of the interatomic bond changes the $\Delta E_v(C) \propto E_b C^{-m}$, accordingly.

Likewise, the second coefficient of the Taylor series corresponds to the frequency of a vibrating dimer or the phonon frequency shift $\Delta\omega$ that is proportional to the curvature of the potential at equilibrium with μ being the reduced mass of the H–O oscillator:

$$\left[\Delta\omega(r)\right]^2 = \left.\frac{d^2 U_{inter}(r)}{\mu dr^2}\right|_{r=d} \propto \frac{E_b}{\mu d^2} \propto Yd \propto (k + k_{Coh}). \tag{5}$$

The $Y \propto E/d^3$ is the elastic modulus and $Yd$ is the stiffness. The $k$ is the force constant of the single vibrating dimer and $k_{Coh}$ is the curvature of the O—O coupling potential for the O:H–O bond[32]. The following formulates the bond length $d$ dependence of the vibration frequency shift $\Delta\omega$ and O1s energy shift $\Delta E_{1s}$:

$$\begin{cases} d/d_b &= C \\ E/E_b &\propto C^{-m} \\ \Delta\omega/\Delta\omega_b &\propto C^{-(1+m/2)} \end{cases} \Rightarrow \begin{cases} Ln(E/E_b)/Ln(C) &= -m \\ Ln(\Delta E_{1s}/\Delta E_{1sb})/Ln(C) &= -m \\ Ln(\Delta\omega/\Delta\omega_b)/Ln(C) &= -(1+m/2) \end{cases} \tag{6}$$

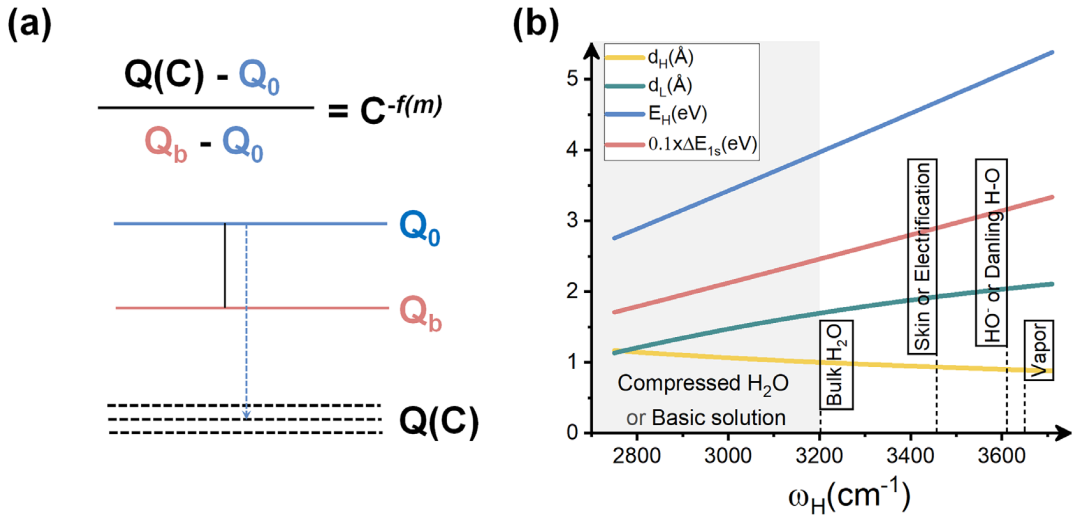

**Figure 2 | Principles and derivatives of H–O bond length-energy-stiffness relaxation.** Illustration of (a) the energy shift of $Q = E_{1s}$ and $\omega_H$ from the reference $Q_0$ to $Q_b$ upon bulk formation and further bond relaxation by perturbation $Q(C) = Q_0 + (Q_b - Q_0)C^{-f(m)}$.[30, 33] The $f(m) = m$ for the $\Delta E_{1s}$ and $(1 + m/2)$ for the $\Delta\omega_H$. (b) Calibration of the $d_H$, $d_L$, $E_H$, and $\Delta E_{1s}$ energy shift as a function of $\omega_H$ with indicated typical scenarios of molecular coordination dependence, as tabulated in **Table 1**.

**Table 1** lists the referential (referenced and given in Figure A2) input ($d_H$, $\omega_H$, $E_{1s}$) values of bulk water, skin, dangling H–O bonds, and perturbation derivatives such as base solutions[26] and water under electrification[34]. The insets in Figure A2a and c show the decomposition of the H–O vibration spectral peak and the X-ray photoelectron spectroscopy (XPS) O1$s$ spectrum of water into the coordination-resolved components of bulk (1.0004, 3200, unseen), skin (uncertain, 3450, 538.10) and the dangling H–O bond (0.90 Å, 3610 cm$^{-1}$, 539.83 eV).

**Table 1 | Initial H–O bond values (referenced) and derivatives.**

| Q | $\omega_H$ (cm$^{-1}$) | $d_H$ (Å) | $d_L$ (Å) | $E_H$ (eV) | $E_{1s}$ (eV) |
|---|---|---|---|---|---|
| $Q_0$ | 1628 ± 1 | 0 | 0 | 0 | 508.2 ± 0.1 |
| $\sigma_r$ | 10$^{-2}$ | - | - | - | 10$^{-5}$ |
| $Q_b$ | 3200[22] | 1.0004[35] | 1.6950[35] | 3.97[25] | 532.80 |
| m = 2.3683; initial input values are indicated with references. | | | | | |
| Skin/electrification | 3450[22, 34] | 0.9350 | 1.921 | 4.66 | 538.10[36] |
| HO$^-$ and Dangling H–O | 3610[22, 26] | 0.8997 | 2.041 | 5.10[37] | 539.83[36] |
| Vapor | 3650[38] | 0.8915 | 2.068 | 5.22 | 540.52 |
| Moon's Water[39] | 3430 (Raman) | 0.9398 | 1.905 | 4.60 | 536.73 |
| | 3480 (IF) | 0.9281 | 1.945 | 4.74 | 537.58 |

Electron and phonon spectroscopy are powerful for collecting information on the energy shift of electrons and phonon or bond stiffness by physical perturbation that causes bond relaxation and electron polarization or chemical reaction of bond reformation and charge transportation, which fingerprints the perturbation to the Hamiltonian and the associated bonding, electronic, and molecular dynamics in the real, time, and energy domains. The spectroscopy sorts bonds of similar force-constant or electrons of close energy regardless of their orientations or locations into a spectral peak whose shape represents the distribution function. Therefore, one can focus on the response of a bond or an electron representing the same sort to a perturbation.

The perturbative differential photoelectron/phonon (PDPS)[33] distills information on the abundance (peak integral), energy (at the highest probability of distribution), and structure order (peak width) transiting from the bulk reference to the conditioned states. The PDPS proceeds by simply subtracting the spectral peak of the bulk reference from the conditioned sample after both peak areas are normalized, as shown in Figure A2a for the radial angle resolved PDPS of water and ice and b for the droplet-size-resolved for water nanodroplets. The former unveiled that the skins of water and ice share the identical H–O bond vibrating at 3450 cm$^{-1}$. The latter resolves the H/D–O bond length of 0.9 Å in the 3.0 Å thick skin of droplets[22].

**Figure 2**a illustrates the principles for the $Q(= \omega_H, E_{1s})$ shifts from the referential $Q_0$ to $Q_b$ upon bulk formation and to $Q(C)$ when subject to bond relaxation under perturbation:

$$\begin{cases} \dfrac{Q(C)-Q_0}{Q(C')-Q_0} = \left(\dfrac{C}{C'}\right)^{-f(m)} & (C \neq C') \quad (a) \\ Q_0 = \dfrac{Q(C)C^{f(m)} - Q(C')C'^{f(m)}}{C^{f(m)} - C'^{f(m)}} & (b) \\ Q(C) = Q_0 + (Q_b - Q_0)C^{-f(m)} & (c) \end{cases} \quad (7)$$

The $Q_0$ is unseen using spectroscopy and the $m$ is unknown, but they can be derived from the known values of $(d, E, \omega)_H$ = (1.0004, 3.97, 3200)$_{bulk}$ and (0.90 Å, 5.1 eV, 3610 cm$^{-1}$)$_{dangling-HB}$. The $m$ = Ln($E/E_b$)/Ln($C$) = Ln(5.1/3.97)/Ln(0.8996) = 2.3683, which gives rise to the referential $\omega_0$ and $E_{1s0}$ in **Table 1**. The $Q_0$ was refined by letting the differentiation of the sum of all the possible $Q_0$ values equal to zero with subscript $l$ and $k$ denoting the afore-mentioned spectral peak components:

$$\frac{d}{dm}\left\{\sum_l \sum_k \frac{Q(C_l)C_l^{f(m)} - Q(C_k)C_k^{f(m)}}{C_l^{f(m)} - C_k^{f(m)}}\right\} = 0. \quad (8)$$

Fine-tuning of the measured $\omega_H(C)$ and $E_{1s}(C)$ values is necessary to meet eq (8) condition, which resulted in the skin mode of 3450 cm$^{-1}$ corresponding to a 0.65% H–O bond contraction or a (0.935+1.921)/2.695-1 = 5.9% O—O expansion, agreeing with that probed using synchrotron electron spectroscopy [40]. **Figure 2**b calibrates the vibration frequency $\omega_H$ dependence of the H–O bond $d_H$-$E_H$-$\Delta E_{1s}$ and the associated $d_L$. Thus, one can read them simultaneously from the references during phonon spectroscopy, as exemplified in the typical situations of bulk water, skin/electrification, dangling bond or OH⁻ solution, and vapor phase, noted in **Figure 2**b.

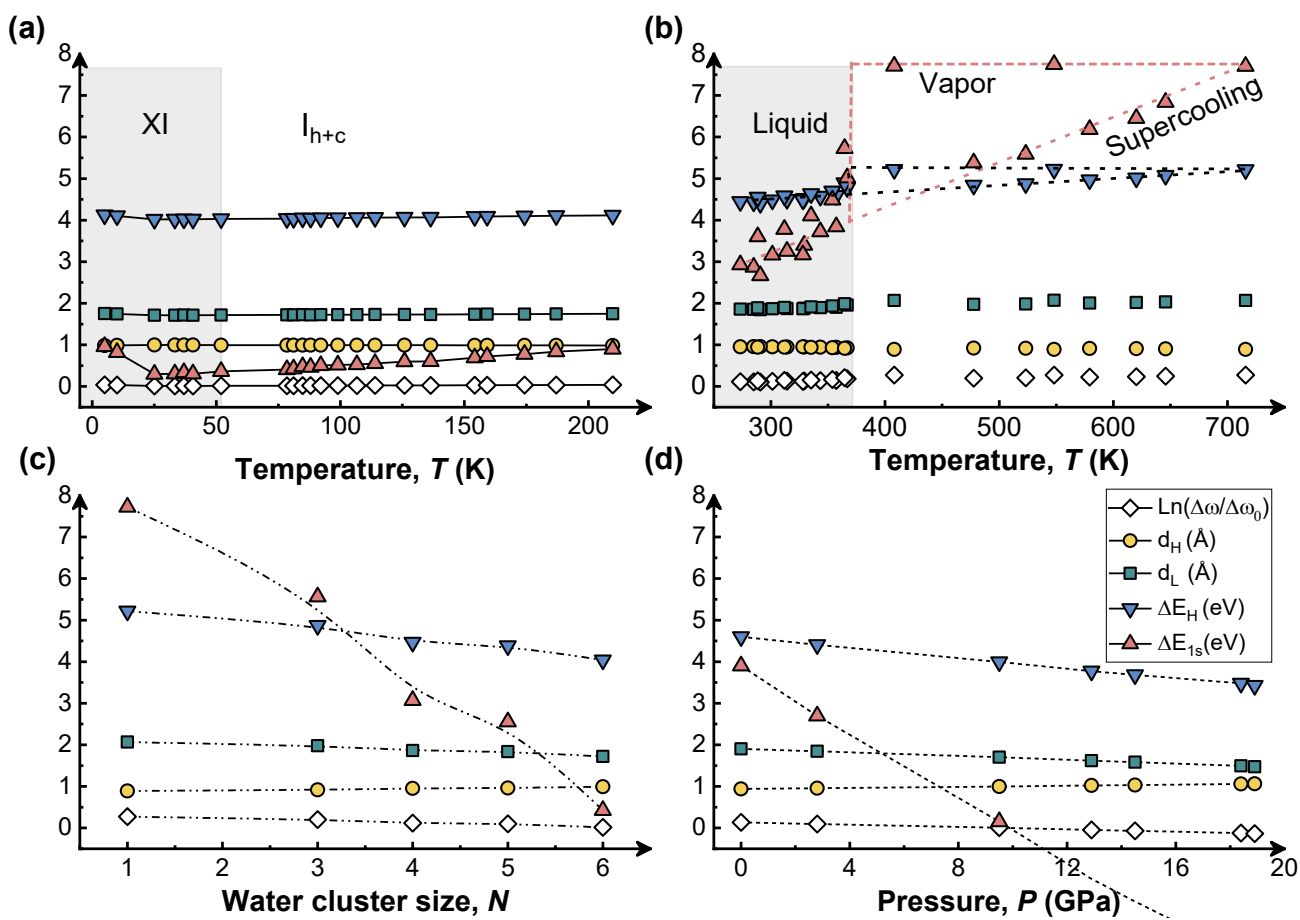

**Figure 3 | Perturbation-resolved H–O bond relaxation.** The H–O bond $d_H$, $d_L$, $E_H$, $\omega_H$, and $\Delta E_{1s}$ relaxation with temperature varying from (a) XI (T < 70 K) and $I_{h+c}$ (70 < T < 275 K) of nanosized ice and (b) Liquid (273 < T < 373 K) and Vapor (T > 275 K) phase of bulk water. (c) Water cluster size N, and (d) pressure dependences.

**Figure 3** exemplifies the determination of the perturbation-resolved H–O bond evolution in the $d_H$, $d_L$, $E_H$, $\omega_H$, and $\Delta E_{1s}$ based on the phonon spectral database (Figure A3) by varying (a, b) temperature[38, 41], (c) water cluster size N[42, 43], and (d) mechanical compression[25]. Water undergoes a phase transition from XI, $I_{h+c}$, Liquid, to Vapor phase, and displays superheating. Molecular undercoordination shortens the $d_H$ and raises the $E_H$ and the $\Delta E_{1s}$; compression does these quantities inversely, consistent with the HBCP regulation for the O:H–O relaxation, shown in Figure A1.

**Table A1** tabulates the H–O bond frequency dependence of the concerned quantities for reference, which is not limited to water and ice. For (Li, Na, K)OH solutions [26], the H–O bond of the OH⁻ vibrates at 3610 cm$^{-1}$, which indicates a 0.90 Å length and 5.10 eV energy for the solute H–O bond; for the $H_2O_2$ solution, the H–O bond vibrates at 3550 cm$^{-1}$ with an estimated 0.91 Å length and 4.94 eV energy. The H–O bond vibrating at 3430(Raman)/3480(IR) cm$^{-1}$ for water on the Moon[39] suggests that the water is either in the form of N ~ 4 ± 1 sized clusters (Figure A3c) or in the state of polarized salt solution (~3450 cm$^{-1}$) though the spectroscopy was conducted at the ambient conditions.

One may also derive the H–O bond flexibility α(q) from eq(6) against a load of q perturbation, such as compression (q = P) and thermal excitation (q = T) of the H–O bond within a certain phase, with dω/dq being detectable (m = 2.3683):

$$\alpha(q) = \frac{dL}{dq} = \frac{dL}{d\omega}\frac{d\omega}{dq} = \frac{d\omega}{dq} \bigg/ \frac{d\omega}{dL}$$
$$= -\frac{d\omega}{dq} \bigg/ \left[\left(1+\frac{m}{2}\right)\Delta\omega_b C^{-(2+m/2)}\right] \quad (9)$$
$$= \frac{-dLn(\Delta\omega)}{\left(2.1841 C^{-3.1841}\right)dq}$$

In summary, incorporating the bond nature index, O—O repulsive coupling, tight-binding approach, and electron and phonon spectroscopy have enabled a referential database from which one can synchronically quantify the relxation and flecxibility of the H–O bond length, energy, vibration frequency, $O_{1s}$ energy, and the O:H distance during spectroscopy for H–O bond-engaged substances. Instead of the long perception of its rigidity and independency, the H–O bond is flexible and sensitive to a perturbation, showing strong cooperativity with the O:H nonbond. The impact of taking the H–O bond flexibility and cooperativity into effect would be profound and tremendous in revisiting the two-body O···H hydrogen bond definition and fresh insight into the H–O bond-involved substances including water and ice.

**Declaration**

No conflicting interest is declared.

**Acknowledgment**

Financial support from the Guangdong Provincial (No. 2024A1515011094(CQ)) and the National Natural Science Foundation of China (No. 12150100 (BW)) are gratefully acknowledged.

**Appendices**

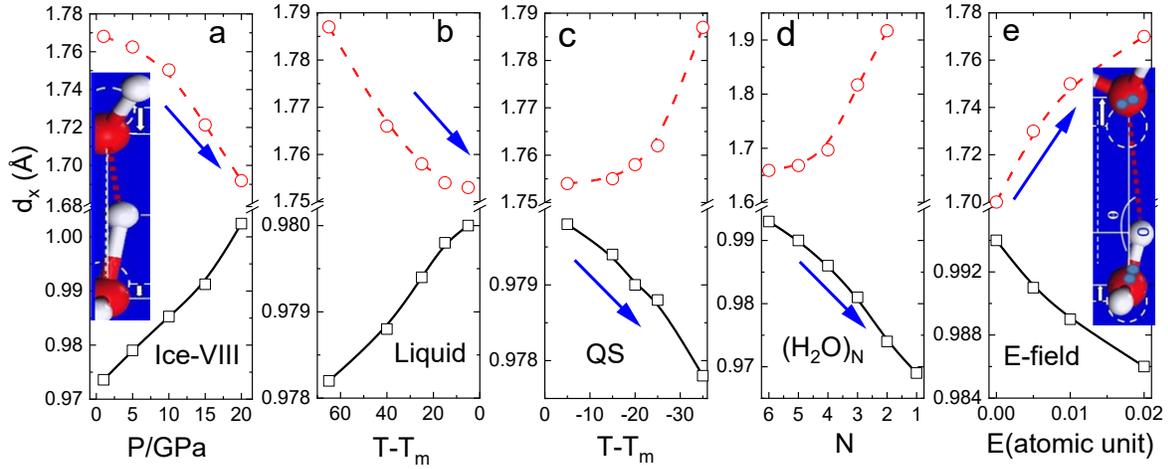

**Figure A1 | O:H−O bond cooperativity and polarizability (HBCP)[22]: (a)** Mechanical compression and **(b)** Liquid cooling shorten and stiffen the weaker O:H nonbond while doing the stronger H−O bond contrastingly(subscript x = H for H−O, and L for O:H interactions); however, **(c)** Quasi-solid (QS) phase cooling, **(d)** molecular undercoordination, and **(e)** electrostatic polarization do the contrast, associated with strong polarization. Insets illustrate that any distance change in the O—O is realized by one part contraction and the other elongation because of the O—O repulsive coupling. Both O atoms shift by different amounts in the same direction concerning the H as the coordination origin.

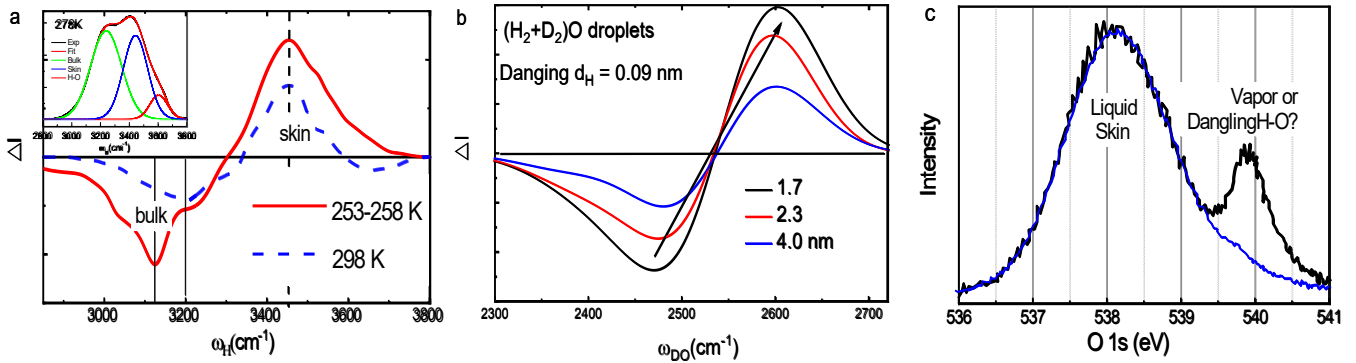

**Figure A2 | Referential data of molecular-CN resolved H/D−O bond vibration frequency and $O_{1s}$ energy.** (a) The $\omega_H$ transits from 3200/3150 cm$^{-1}$ for the 298/255 K water/ice [44] to their common skin

mode of 3450 cm$^{-1}$. (b) Droplet size reduction shifts the $\omega_D$ from its bulk below 2550 cm$^{-1}$ to the skin mode above [45]. (c) The XPS O 1s energy shifts from the bulk value of unseen to 538.1 eV for the liquid skin and 539.9 eV for the monomer of the gaseous phase (should be a dangling bond, instead) [46]. Inset a decomposes the H–O spectral peak into the coordination resolved components of bulk (3200), skin (3450), and the dangling H–O bond (3610 cm$^{-1}$).

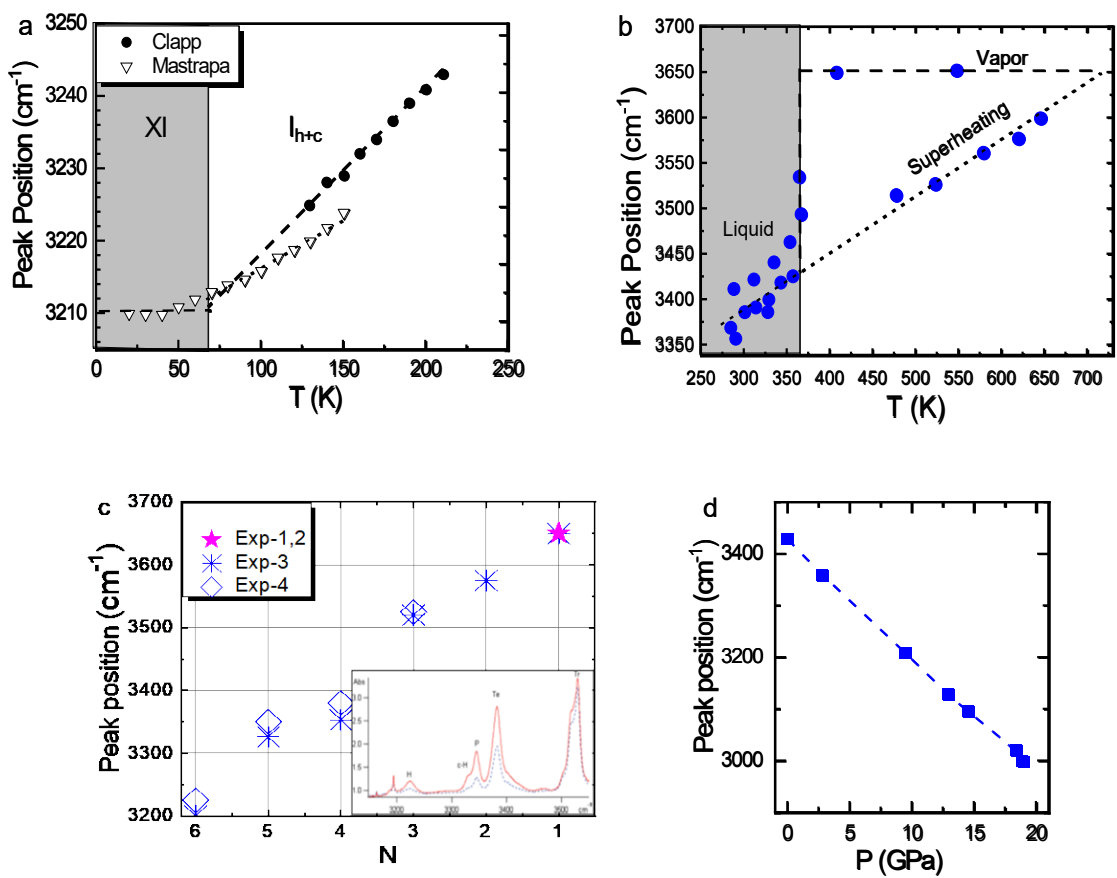

**Figure A3 | Referential data of perturbation-resolved H–O bond vibration frequency shift.** Thermal $\omega_H$ shifting for (a) nanosized ice at low temperatures [41, 47] showing the H–O bond silence because the specific heat approaches zero[31] and (b) bulk water in the 273 - 573 K temperature range showing Liquid-Vapor phase transition and superheating[38] (reproduced with permission from [38, 41, 47] and references therein). Cluster size dependence of (c) the $\omega$ relaxation in comparison with (b) measurements (scattered data)[38, 42, 43] for (H$_2$O)$_N$ clusters. (d) Typical pressure-resolved $\omega_H$ and $\omega_L$ shift [25].

**Table A1 | Referential database of the $\omega_H$ dependence of the H–O bond identities and O:H distance.**

| $\omega_H(cm^{-1})$ | $Ln(\Delta\omega/(\Delta\omega_b)$ | $d_H$ (Å) | $d_L$ (Å) | $E_H$ (eV) | $\Delta E_{1s}$ (eV) | Notes |
|---|---|---|---|---|---|---|
| 2750 | -0.337 | 1.167 | 1.134 | 2.75 | 17.07 | Compressed O:H–O bond (pressure[22] & O:⇔:O repulsion in base OH- and $H_2O_2$ solutions[26]) |
| 2770 | -0.320 | 1.158 | 1.163 | 2.81 | 17.40 | |
| 2790 | -0.302 | 1.149 | 1.192 | 2.86 | 17.73 | |
| 2810 | -0.285 | 1.140 | 1.221 | 2.91 | 18.06 | |
| 2830 | -0.268 | 1.131 | 1.249 | 2.97 | 18.39 | |
| 2850 | -0.252 | 1.123 | 1.277 | 3.02 | 18.72 | |
| 2870 | -0.236 | 1.114 | 1.304 | 3.07 | 19.05 | |
| 2890 | -0.220 | 1.106 | 1.331 | 3.13 | 19.39 | |
| 2910 | -0.204 | 1.098 | 1.358 | 3.18 | 19.72 | |
| 2930 | -0.188 | 1.091 | 1.384 | 3.24 | 20.05 | |
| 2950 | -0.173 | 1.083 | 1.409 | 3.29 | 20.39 | |
| 2970 | -0.158 | 1.076 | 1.434 | 3.34 | 20.72 | |
| 2990 | -0.143 | 1.068 | 1.459 | 3.40 | 21.06 | |
| 3010 | -0.129 | 1.061 | 1.484 | 3.45 | 21.39 | |
| 3030 | -0.114 | 1.054 | 1.508 | 3.51 | 21.73 | |
| 3050 | -0.100 | 1.047 | 1.531 | 3.56 | 22.07 | |
| 3070 | -0.086 | 1.041 | 1.554 | 3.62 | 22.40 | |
| 3090 | -0.073 | 1.034 | 1.577 | 3.67 | 22.74 | |
| 3110 | -0.059 | 1.028 | 1.599 | 3.72 | 23.08 | |
| 3130 | -0.046 | 1.021 | 1.621 | 3.78 | 23.41 | |
| 3150 | -0.032 | 1.015 | 1.643 | 3.83 | 23.75 | |
| 3170 | -0.019 | 1.009 | 1.664 | 3.89 | 24.09 | |
| **3190** | **-0.006** | **1.003** | **1.684** | **3.94** | **24.43** | Bulk $H_2O$[22] |
| **3210** | **0.006** | **0.997** | **1.705** | **4.00** | **24.77** | |
| 3230 | 0.019 | 0.992 | 1.725 | 4.05 | 25.11 | |
| 3250 | 0.031 | 0.986 | 1.744 | 4.11 | 25.45 | |
| 3270 | 0.044 | 0.981 | 1.764 | 4.16 | 25.79 | |
| 3290 | 0.056 | 0.975 | 1.782 | 4.22 | 26.13 | |
| 3310 | 0.068 | 0.970 | 1.801 | 4.27 | 26.47 | |
| 3330 | 0.079 | 0.965 | 1.819 | 4.33 | 26.81 | |
| 3350 | 0.091 | 0.960 | 1.837 | 4.38 | 27.16 | |
| 3370 | 0.103 | 0.954 | 1.854 | 4.44 | 27.50 | |
| 3390 | 0.114 | 0.949 | 1.872 | 4.49 | 27.84 | |
| 3410 | 0.125 | 0.945 | 1.889 | 4.55 | 28.18 | |
| 3430 | 0.137 | 0.940 | 1.905 | 4.60 | 28.53 | |
| **3450** | **0.148** | **0.935** | **1.921** | **4.66** | **28.87** | Water ice skin[22]/Ionic solutions[26]/water bridge[48] |
| 3470 | 0.159 | 0.930 | 1.937 | 4.71 | 29.21 | |
| 3490 | 0.169 | 0.926 | 1.953 | 4.77 | 29.56 | |
| 3510 | 0.180 | 0.921 | 1.968 | 4.83 | 29.90 | |
| 3530 | 0.191 | 0.917 | 1.983 | 4.88 | 30.25 | |

| 3550 | 0.201 | 0.912 | 1.998 | 4.94 | 30.59 | |
| 3570 | 0.211 | 0.908 | 2.013 | 4.99 | 30.94 | |
| 3590 | 0.222 | 0.904 | 2.027 | 5.05 | 31.28 | |
| **3610** | **0.232** | **0.900** | **2.041** | **5.10** | **31.63** | Dangling H–O bond[22] |
| 3630 | 0.242 | 0.896 | 2.055 | 5.16 | 31.97 | |
| **3650** | **0.252** | **0.891** | **2.068** | **5.22** | **32.32** | Vapor[37, 38] |
| 3670 | 0.262 | 0.887 | 2.082 | 5.27 | 32.67 | |
| 3690 | 0.271 | 0.884 | 2.095 | 5.33 | 33.01 | |
| 3710 | 0.281 | 0.880 | 2.108 | 5.38 | 33.36 | |

**References**

1. Arunan, E.; Desiraju, G. R.; Klein, R. A.; Sadlej, J.; Scheiner, S.; Alkorta, I.; Clary, D. C.; Crabtree, R. H.; Dannenberg, J. J.; Hobza, P.; Kjaergaard, H. G.; Legon, A. C.; Mennucci, B.; Nesbitt, D. J., Defining the hydrogen bond: An account (IUPAC Technical Report). *Pure Appl. Chem.* **2011,** *83* (8), 1619-1636.
2. Lennard-Jones, J. E., On the forces between atoms and ions. *Proceedings of the Royal Society of London. Series A, Containing Papers of a Mathematical and Physical Character* **1925,** *109* (752), 584-597.
3. Bernal, J. D.; Fowler, R. H., A Theory of Water and Ionic Solution, with Particular Reference to Hydrogen and Hydroxyl Ions. *J. Chem. Phys.* **1933,** *1* (8), 515-548.
4. Berendsen, H. J.; Postma, J. P.; van Gunsteren, W. F.; Hermans, J. In *Interaction models for water in relation to protein hydration*, Intermolecular forces: proceedings of the fourteenth Jerusalem symposium on quantum chemistry and biochemistry held in jerusalem, israel, april 13–16, 1981, Springer: 1981; pp 331-342.
5. Jorgensen, W. L.; Chandrasekhar, J.; Madura, J. D.; Impey, R. W.; Klein, M. L., Comparison of simple potential functions for simulating liquid water. *J. Chem. Phys.* **1983,** *79* (2), 926-935.
6. Jorgensen, W. L.; Madura, J. D., Temperature and size dependence for Monte Carlo simulations of TIP4P water. *Mol. Phys.* **1985,** *56* (6), 1381-1392.
7. Mahoney, M. W.; Jorgensen, W. L., A five-site model for liquid water and the reproduction of the density anomaly by rigid, nonpolarizable potential functions. *J. Chem. Phys.* **2000,** *112* (20), 8910-8922.
8. Caldwell, J. W.; Kollman, P. A., Structure and properties of neat liquids using nonadditive molecular dynamics: water, methanol, and N-methylacetamide. *J. Chem. Phys.* **1995,** *99* (16), 6208-6219.
9. Berendsen, H. J.; Grigera, J.-R.; Straatsma, T. P., The missing term in effective pair potentials. *J. Phys. Chem.* **1987,** *91* (24), 6269-6271.
10. Ponder, J. W.; Case, D. A., Force fields for protein simulations. *Adv. Protein Chem.* **2003,** *66*, 27-85.
11. McBride, C.; Vega, C.; Noya, E. G.; Ramírez, R.; Sesé, L. M., Quantum contributions in the ice phases: The path to a new empirical model for water—TIP4PQ/2005. *J. Chem. Phys.* **2009,** *131* (2), 024506.
12. Tang, F.; Li, Z.; Zhang, C.; Louie, S. G.; Car, R.; Qiu, D. Y.; Wu, X., Many-body effects in the X-ray absorption spectra of liquid water. *Proceedings of the National Academy of Sciences* **2022,** *119* (20), e2201258119.
13. Niu, H.; Yang, Y. I.; Parrinello, M., Temperature dependence of homogeneous nucleation in ice. *Phys. Rev. Lett.* **2019,** *122* (24), 245501.
14. Slater, B.; Michaelides, A., Surface premelting of water ice. *Nature Reviews Chemistry* **2019,** *3* (3), 172-188.
15. Goga, N.; Mayrhofer, L.; Tranca, I.; Nedea, S.; Heijmans, K.; Ponnuchamy, V.; Vasilateanu, A., A review of recent developments in molecular dynamics simulations of the photoelectrochemical water splitting process. *Catalysts* **2021,** *11* (7), 807.
16. Yang, J.; Dettori, R.; Nunes, J. P. F.; List, N. H.; Biasin, E.; Centurion, M.; Chen, Z.; Cordones, A. A.; Deponte, D. P.; Heinz, T. F.; Kozina, M. E.; Ledbetter, K.; Lin, M.-F.; Lindenberg, A. M.; Mo, M.; Nilsson, A.; Shen, X.; Wolf, T. J. A.; Donadio, D.; Gaffney, K. J.; Martinez, T. J.; Wang, X., Direct observation of ultrafast hydrogen bond strengthening in liquid water. *Nature* **2021,** *596* (7873), 531-535.
17. Faraday, M., Note on Regelation. *Proceedings of the Royal Society of London* **1860,** *10*, 440-450.
18. Mpemba, E. B.; Osborne, D. G., Cool? *Phys. Educ.* **1979,** *14*, 410-413.
19. Mallamace, F.; Branca, C.; Broccio, M.; Corsaro, C.; Mou, C. Y.; Chen, S. H., The anomalous behavior of the density of water in the range 30 K < T < 373 K. *Proc. Natl. Acad. Sci. U.S.A.* **2007,** *104* (47), 18387-91.

20. Montero de Hijes, P.; Dellago, C.; Jinnouchi, R.; Kresse, G., Density isobar of water and melting temperature of ice: Assessing common density functionals. *J. Chem. Phys.* **2024,** *161* (13).
21. Montero de Hijes, P.; Dellago, C.; Jinnouchi, R.; Schmiedmayer, B.; Kresse, G., Comparing machine learning potentials for water: Kernel-based regression and Behler–Parrinello neural networks. *J. Chem. Phys.* **2024,** *160* (11).
22. Sun, C. Q.; Huang, Y.; Zhang, X.; Ma, Z.; Wang, B., The physics behind water irregularity. *Physics Reports* **2023,** *998*, 1-68.
23. Sun, C. Q., Oxidation electronics: bond-band-barrier correlation and its applications. *Prog. Mater Sci.* **2003,** *48* (6), 521-685.
24. Goncharov, A. F.; Struzhkin, V. V.; Mao, H.-k.; Hemley, R. J., Raman spectroscopy of dense H 2 O and the transition to symmetric hydrogen bonds. *Phys. Rev. Lett.* **1999,** *83* (10), 1998.
25. Sun, C. Q.; Zhang, X.; Zheng, W. T., Hidden force opposing ice compression. *Chem Sci* **2012,** *3*, 1455-1460.
26. Sun, C. Q., Aqueous charge injection: solvation bonding dynamics, molecular nonbond interactions, and extraordinary solute capabilities. *Int. Rev. Phys. Chem.* **2018,** *37* (3-4), 363-558.
27. Wang, J.; Zeng, Y.; Zheng, Z.; Zhang, L.; Wang, B.; Yang, Y.; Sun, C. Q., Discriminative Mechanical and Thermal Response of the H-N Bonds for the Energetic LLM-105 Molecular Assembly. *J. Phys. Chem. Lett.* **2023,** *14*, 8555−8562.
28. Zhang, L.; Yao, C.; Yu, Y.; Jiang, S.-L.; Sun, C. Q.; Chen, J., Stabilization of the Dual-Aromatic Cyclo-N5⁻ Anion by Acidic Entrapment. *Journal of physical chemistry letters* **2019,** *10*, 2378–2385.
29. Sun, C. Q., Size dependence of nanostructures: Impact of bond order deficiency. *Prog. Solid State Chem.* **2007,** *35*, 1-159.
30. Liu, X. J.; Zhang, X.; Bo, M. L.; Li, L.; Nie, Y. G.; Tian, H.; Sun, Y.; Xu, S.; Wang, Y.; Zheng, W.; Sun, C. Q., Coordination-resolved electron spectrometrics. *Chem. Rev.* **2015,** *115* (14), 6746-6810.
31. Omar, M. A., *Elementary Solid State Physics: Principles and Applications*. Addison-Wesley: New York, 1993.
32. Huang, Y.; Ma, Z.; Zhang, X.; Zhou, G.; Zhou, Y.; Sun, C. Q., Hydrogen Bond Asymmetric Local Potentials in Compressed Ice. *J. Phys. Chem. B* **2013,** *117* (43), 13639-13645.
33. Yang, X.; Peng, C.; Li, L.; Bo, M.; Sun, Y.; Huang, Y.; Sun, C. Q., Multifield-resolved phonon spectrometrics: structured crystals and liquids. *Prog. Solid State Chem.* **2019,** *55*, 20-66.
34. Fuchs, E. C.; Sammer, M.; Wexler, A. D.; Kuntke, P.; Woisetschläger, J., A floating water bridge produces water with excess charge. *J. Phys. D: Appl. Phys.* **2016,** *49* (12), 125502.
35. Huang, Y.; Zhang, X.; Ma, Z.; Zhou, Y.; Zhou, J.; Zheng, W.; Sun, C. Q., Size, separation, structure order, and mass density of molecules packing in water and ice. *Sci Rep* **2013,** *3*, 3005.
36. Nishizawa, K.; Kurahashi, N.; Sekiguchi, K.; Mizuno, T.; Ogi, Y.; Horio, T.; Oura, M.; Kosugi, N.; Suzuki, T., High-resolution soft X-ray photoelectron spectroscopy of liquid water. *Physical Chemistry Chemical Physics* **2011,** *13*, 413-417.
37. Harich, S. A.; Hwang, D. W. H.; Yang, X.; Lin, J. J.; Yang, X.; Dixon, R. N., Photodissociation of H2O at 121.6 nm: A state-to-state dynamical picture. *J. Chem. Phys.* **2000,** *113* (22), 10073-10090.
38. Cross, P. C.; Burnham, J.; Leighton, P. A., The Raman spectrum and the structure of water. *Journal of the American Chemical Society* **1937,** *59*, 1134-1147.
39. Jin, S.; Hao, M.; Guo, Z.; Yin, B.; Ma, Y.; Deng, L.; Chen, X.; Song, Y.; Cao, C.; Chai, C.; Wei, Q.; Ma, Y.; Guo, J.; Chen, X., Evidence of a hydrated mineral enriched in water and ammonium molecules in the Chang'e-5 lunar sample. *Nature Astronomy* **2024,** *8* (9), 1127-1137.
40. Wilson, K. R.; Schaller, R. D.; Co, D. T.; Saykally, R. J.; Rude, B. S.; Catalano, T.; Bozek, J. D., Surface relaxation in liquid water and methanol studied by x-ray absorption spectroscopy. *J. Chem. Phys.* **2002,** *117* (16), 7738-7744.
41. Medcraft, C.; McNaughton, D.; Thompson, C. D.; Appadoo, D.; Bauerecker, S.; Robertson, E. G., Size and Temperature Dependence in the Far-Ir Spectra of Water Ice Particles. *The Astrophysical Journal* **2012,** *758* (1), 17.
42. Hirabayashi, S.; Yamada, K. M. T., Infrared spectra and structure of water clusters trapped in argon and krypton matrices. *J. Mol. Struct.* **2006,** *795* (1-3), 78-83.
43. Ceponkus, J.; Uvdal, P.; Nelander, B., Water tetramer, pentamer, and hexamer in inert matrices. *J. Phys. Chem. A* **2012,** *116* (20), 4842-50.
44. Kahan, T. F.; Reid, J. P.; Donaldson, D. J., Spectroscopic probes of the quasi-liquid layer on ice. *J. Phys. Chem. A* **2007,** *111* (43), 11006-11012.
45. Park, S.; Moilanen, D. E.; Fayer, M. D., Water Dynamics: The Effects of Ions and Nanoconfinement. *J. Phys. Chem. B* **2008,** *112* (17), 5279-5290.
46. Winter, B.; Aziz, E. F.; Hergenhahn, U.; Faubel, M.; Hertel, I. V., Hydrogen bonds in liquid water studied by photoelectron spectroscopy. *J. Chem. Phys.* **2007,** *126* (12), 124504.

47. Medcraft, C.; McNaughton, D.; Thompson, C. D.; Appadoo, D. R. T.; Bauerecker, S.; Robertson, E. G., Water ice nanoparticles: size and temperature effects on the mid-infrared spectrum. *Physical Chemistry Chemical Physics* **2013,** *15* (10), 3630-3639.
48. Ponterio, R.; Pochylski, M.; Aliotta, F.; Vasi, C.; Fontanella, M.; Saija, F., Raman scattering measurements on a floating water bridge. *J. Phys. D: Appl. Phys.* **2010,** *43* (17), 175405.